\documentstyle[aasms4,12pt]{article}
\def\lax    {\ifmmode{_<\atop^{\sim}}\else{${_<\atop^{\sim}}$}\fi}
\def\gax    {\ifmmode{_>\atop^{\sim}}\else{${_>\atop^{\sim}}$}\fi}

\def\gtorder{\mathrel{\raise.3ex\hbox{$>$}\mkern-14mu
	     \lower0.6ex\hbox{$\sim$}}}
\def\ltorder{\mathrel{\raise.3ex\hbox{$<$}\mkern-14mu
	     \lower0.6ex\hbox{$\sim$}}}
\def\gsim{\mathrel{\raise.3ex\hbox{$>$}\mkern-14mu
	     \lower0.6ex\hbox{$\sim$}}}
\def\lsim{\mathrel{\raise.3ex\hbox{$<$}\mkern-14mu
	     \lower0.6ex\hbox{$\sim$}}}

\begin{document}


\title{Probing the Structure of Accreting Compact Sources Through 
X-Ray Time Lags and Spectra}

\author{Xin-Min Hua\altaffilmark{1}, Demosthenes Kazanas } 
\affil{LHEA, NASA/GSFC Code 661, Greenbelt, MD 20771}
\medskip
\centerline{and}
\author{Wei~Cui}
\affil{Center for Space Research, MIT, Cambridge, MA 02139, USA}

\altaffiltext{1}{Universities Space Research Association}


\begin{abstract}

We exhibit, by compiling all data sets we can acquire, that 
the Fourier frequency dependent hard X-ray lags, first observed in
the analysis of aperiodic variability of the light curves of the black
hole candidate Cygnus X-1, appear to be a property shared by 
several other accreting black hole candidate sources and also by the
different spectral states of this source. We then present both 
analytic and numerical models of these time lags resulting by the 
process of Comptonization in a variety of hot electron configurations. 
We argue that under the assumption that the observed spectra are due to 
Comptonization, the dependence of the lags on the Fourier period 
provides a means for mapping the spatial density profile of the hot 
electron plasma, while the period at which the lags eventually 
level--off provides an estimate of the size of the scattering cloud. 
We further examine the influence of the location and spatial extent 
of the soft photon source on the form of the resulting lags for a 
variety of configurations; we conclude that the study of the X-ray 
hard lags can provide clues about these parameters of the 
Comptonization process too. Fits of the existing data with our models
indicate that the size of the Comptonizing clouds are quite large 
in extent ($\sim$ 1 light second) with inferred radial density 
profiles which are in many instances inconsistent with those of 
the standard dynamical models, while the extent of the source
of soft photons appears to be much smaller than those of the hot 
electrons by roughly two orders of magnitude and its location 
consistent with the center of the hot electron corona.

\end{abstract}


\section{INTRODUCTION}

The study of the physics of accretion onto compact objects (neutron stars
and black holes) whether in galactic (X-ray binaries) or extragalactic
systems (Active Galactic Nuclei) involves length scales much too small to
be resolved by current technology or that of the foreseeable future. As 
such, this study is conducted mainly through the theoretical interpretation 
of spectral and temporal observations of these systems, much in the way
that the study of spectroscopic binaries has been used to deduce the 
properties of the binary system members and the elements of their
orbits. In this endeavor, the first line of attack in uncovering the
physical properties of these systems is the analysis of their spectra. 
Thus there exists a large body of spectral observations, accumulated 
over the past several decades, indicating that the spectra of this class 
of objects and in particular the Black Hole Candidate (BHC) sources, 
can be fitted very well by those resulting from the Comptonization of 
soft photons by hot ($T_e \sim 10^9$ K) electrons; the latter are 
``naturally" expected to be present in these sources, a result of the 
dissipation of the accretion kinetic energy within an accretion disk. 
It is thus generally agreed upon that Comptonization is the process 
responsible for the production of the high energy ($\gsim 2-100$  keV)
radiation in these sources. For this reason, this process has been studied
extensively both analytically and numerically over the past couple of 
decades (see e.g. Sunyaev \& Titarchuk 1980; Titarchuk 1994; Hua \& 
Titarchuk 1995). Thus, while the issue of the detailed dynamics of 
accretion onto the compact object is still not resolved, there has been
ample of spectral evidence corroborating the presence of a population 
of hot electrons, located presumably (for energetic reasons) in its 
vicinity. Spectral modeling and fits to observations have subsequently 
been employed as a means of constraining or even determining the 
dynamics of accretion flows onto the compact  object.  

It is well known, however, that the Comptonization spectra cannot 
provide, in and of themselves, any information about the size of the 
scattering plasma, because they depend (for optically thin plasmas) only
on the product of the electron temperature and the plasma Thomson depth.
Therefore, they cannot provide any clues about the dynamics of accretion 
of the hot gas onto the compact object, which require the knowledge of 
the density and velocity as a function radius. To determine the dynamics 
of accretion, one needs, in addition to the spectra, time 
variability information. It is thought, however, that such information 
may not be terribly relevant, because it is generally accepted that the 
X-ray emission originates at the smallest radii of the accreting flow 
and as such, time variability would simply reflect the dynamical or 
scattering time scales of the emission region, of order of msec for 
galactic accreting sources and $10^5 -10^7$  times longer for AGN
(one should bear in mind however, that for black holes masses as 
large as expected in AGN, the emission associated with the region 
closest to the black hole may in fact be that of the UV rather than
X-ray part of the spectrum, without that invalidating the arguments
concerning their X-ray emission, which is also generally accepted to
be due to Comptonization).
  
Variability studies, however, have provided inconclusive, at best, 
results: The simplest and most frequently used variability measure, 
the Power Spectral Density (hereafter PSD), has the following generic,
counter-intuitive form (see e.g Miyamoto et al. 1991; Belloni et 
al. 1996; Grove et al. 1997, Cui et al. 1997b)
\begin{equation}
\vert F(\nu) \vert^2 \propto \nu^{-s} ~~~~{\rm with} 
\cases {s \simeq 0 &for $\nu < \nu_c$ \cr
s \gsim 1 &for $\nu_b >\nu > \nu_c $ \cr
s \simeq 2 &for $\nu > \nu_b $ \cr }, 
\end{equation}
where $\nu_c \simeq 0.3$ Hz and $\nu_b \gsim 3 - 10$ Hz. The surprising 
feature of PSDs with the above form is that most of the variability 
power resides around $\nu \simeq \nu_c$, a frequency $3-4$ orders of 
magnitude lower than the kHz range characteristic of the presumed 
dynamics of infall onto the compact object. There is a fair amount 
of variability power between $\nu_c$ and $\nu_b \simeq 10$ Hz, with 
the PSD slope being flatter than the value corresponding to simple 
exponential shots ($s=2$) and consistent with the slope of ``flicker 
noise" ($s \simeq 1$), a fact thought to be of significance for 
understanding the underlying dynamics. However, it is our opinion that
of greater significance is the marked absence of variability at the 
anticipated kHz range of frequencies, in view of the fact that the 
apparently thin emission requires infall velocities close to that of 
free--fall; this latter fact implies, then, that the flows near the 
compact object should have a significant turbulent component 
associated with the viscosity necessary to yield the inferred large 
infall velocities; we find of interest that the presence of this
turbulence is not apparent in the observed X-ray variability, as 
measured by the PSD. It is worth noting that a
similar lack of variability at the expected frequencies was also found for 
AGN by Tennant \& Mushotzky (1983) in the HEAO-1 database. The PSDs of 
AGN X-ray light curves, though not as firmly established as those of 
galactic BHC, appear nonetheless to indicate a similar lack of 
variability at the corresponding anticipated frequency range ($\sim 
10^{-2.5}$ Hz) with most of the power at significantly lower 
frequencies ($\sim 10^{-6}$ Hz; Green et al. 1993).

This discrepancy between the expected and observed variability, is
generally attributed to a modulation of the accretion rate 
onto the compact object, thus deferring an account of the PSD form
to a more thorough understanding of the dynamics of accretion.
However, the Comptonization process, thought to be responsible for 
the high energy emission, offers an alternative measure of 
variability  more refined than the PSD (see e.g. van der Klis et 
al. 1987; Wijers, van Paradijs \& Lewin 1987): In this process, 
the energy of the scattered  soft photons increases, on the average, 
with their residence time in the scattering medium; as a result, 
the hard photon light curves lag with respect to those of softer 
photons by amounts which are roughly proportional to the photon 
scattering time and the logarithm of the photon energy ratio. Thus, 
observations of these time lags provide a measure of the local 
electron density, or in an optically thin medium ($\tau_0 \sim 1$)
an estimate of its size. Such an estimate is inaccessible to 
analysis restricted only to spectral fits or simply to measurements 
of the PSDs. Therefore, if the high energy 
radiation is produced in the vicinity of the compact object and the 
observed PSDs are due to a modulation of the accretion rate by a 
hitherto unknown process, as most models contend (see e.g. Chen \& Taam
1995; Takeuchi et al. 1995), these lags should be 
roughly of the order $\sim$ msec or shorter, the electron scattering 
time near the compact object, independent of the modulation of overall 
fluctuations of the light curve.

Motivated by the above considerations and the apparent discrepancy 
between the observed and expected variability time scales of BHC, 
we have recently embarked in the study of the timing properties 
of inhomogeneous, hot electron clouds of significant spatial extent;
our goal in this effort was to provide models for the light curves
and other timing properties (PSD, lags) of BHC which would attribute
some physical significance to the features of their PSDs discussed 
above. Thus, in Kazanas, Hua \& Titarchuk (1997; hereafter
KHT) we have studied the time response of such extended sources and 
indicated that the observed PSD could be understood as resulting from 
Comptonization of soft photons by hot electron configurations which 
extend over several orders of magnitude in radius, identifying 
the break at $\nu_c$ in the PSD with the ``outer edge" of the 
scattering cloud. In Hua, Kazanas \& Titarchuk (1997; hereafter 
HKT) we examined the Coherence Function (hereafter CF; Vaughan 
\& Nowak 1997) associated with such spatially extended hot electron 
configurations and indicated that, due to the linearity of the 
Comptonization process, these configurations are expected to exhibit 
a high degree of coherence, as well as long hard X-ray lags, in 
agreement with observations. More recently, Kazanas \& Hua (1998) 
have produced model light curves associated with these configurations 
and studied their properties in the time domain, using moments of 
various orders which they then correlate with other observables.

In the present work we focus our attention to a more detailed study of 
the lags between different bands of the X-ray light curves. In \S 2 we 
collect all the lag and spectral  observations associated with the light 
curves of BHC which we were able to obtain in the literature; we 
indicate that these lags have a form which appears to be very similar 
among these different sources and which does not correlate in an 
obvious way with their well-studied spectral states, and we provide 
fits to the observed lags using our earlier models. In \S 3 we 
provide a heuristic account of the form of the observed lags and 
their dependence on the density distribution of the hot Comptonizing 
electrons, as well as models which allow one to compute the form 
of lags analytically thus making transparent their origin.  In \S 4 
we demonstrate, by direct numerical simulation,  the dependence of 
the lag form on both the spatial distribution of Comptonizing electrons 
and the location of the source of soft photons, and provide the form 
of the lags associated with a variety of such configurations.  
Finally, in \S 5 the results are summarized and conclusions are drawn. 

\section{The Lag Observations}

As indicated in the introduction, the need for models with timing 
properties which are in agreement with the observations of the PSDs and 
the hard X-ray lags of BHC  motivated KHT to study the timing 
properties of the Comptonization process by {\it inhomogeneous, extended}, 
hot electron configurations of the form
\begin{equation}
n(r) = \cases  {n_i &for $r \le r_1$ \cr n_1 (r_1/r)^{p} &
for $r_2 > r > r_1$ \cr}
\label{density} 
\end{equation}
where the power index $p$ is a free parameter; $r$ is the radial distance 
from the center of the spherical corona; $r_1$ and $r_2$ are the radii 
of its inner and outer edges respectively. Note that the density of the 
central uniform core, $n_i$, does not necessarily join smoothly to that 
of the extended corona. It is apparent that the density profile of 
Equation (\ref{density}) allows Compton scattering to take place over a 
wide range of densities, thereby introducing time lags over a similar 
range of time scales and at the same time producing variability over a 
similar frequency range, thus providing a straight forward account 
and physical significance to the observed PSDs (KHT).

In figure 1 we present the time lags associated with the Ginga 
observations of Cyg X-1 in August 1987, when the source was 
apparently in its {\it hard} state, as reported by Miyamoto et al. 
(1988). On the same figure we also plot the times lags associated 
with a variety of electron configurations of the general form of Eq. 
(\ref{density}), as computed using the Monte Carlo code of Hua (1997). 
These were computed by injecting the soft photons at $r=0$ from 
a black body distribution of $kT_r = 0.2$ keV to an electron cloud with 
temperature $kT_e = 100$ keV and Thomson depth $\tau_0 =1$. 
The escaping photons were collected according to their arrival time 
to the observer as well as their energy. The energy bins used were 
$15.8 - 24.4$ and $1.2 - 5.75$ keV, in order to be directly 
compared to the observational data. In each energy bin, the photons 
were collected into 4096 time bins over 16 seconds, each $1/256$ seconds 
in length. The light curves so obtained were then used to calculate the 
time lags of the emission in the higher energy bin with respect to that 
in the lower one. 

The time lags resulting from Comptonization of the soft photons a 
uniform corona (dotted curve) were obtained by further setting its 
density to $n = 10^{16}$ cm$^{-3}$. It is seen that the lags are 
constant $\approx 2$ ms down to a period $\sim 0.005$ second and then 
decrease at smaller periods. This is because for emission 
from such a corona, the hard X-ray lags are due to scatterings in a 
region with a mean free time of the order $0.3/n\sigma_T c \simeq 1.5$ 
ms (Hua \& Titarchuk, 1996). The magnitude of the time lag reflects 
this characteristic time. The dotted-dashed curve represents the lag 
form for a uniform source of the above density obtained using the
analytic formula of Payne (1980). The difference in the lag normalization 
is indicative of the corrections introduced by using the Klein-Nishina,
as opposed to the Thomson cross section used in the analytic formula, 
in  the scattering time expression.
The solid line represents the lags corresponding to a corona with 
$p=1$, $n_i = n_1 \sim 10^{16}$ cm$^{-3}$, $r_1 = 10^{-3}$ light
seconds, $r_2 = 10^3 r_1$  and electron temperature 100 keV. 
This curve has a linear dependence on the Fourier period ranging from 
$P=0.03$ second up to $\sim 3$ second. There is a cutoff at the periods 
below $P=0.03$ second due to the finite time resolution of our 
calculation. At periods $P \gsim 4$ seconds the curve levels--off,
indicating that the time lag of hard X-ray reaches its maximum of 
$\sim 0.1$ second. The wide range of the lags reflects the fact that
the scatterings take place in a region with densities ranging from 
$\sim 10^{16}$ to $10^{12}$ cm$^{-3}$. The time lags resulting from 
the corona with $p=3/2$ density profile are represented by the dashed 
curve. It is seen that dependence of lag on period is weaker and the 
maximum time lag is only $\sim 10^{-2}$ second, indicating that 
although the corona extends to a radius of $\sim 1$ light second, 
there is virtually no photon scattering beyond 
a radius $\sim 10^{-2}$ light second because of the
steep density gradient. A calculation based on the 
density profile of Equation (2) indicates that the optical depth of the 
corona for $r > 10^{-2}$ light second is $\sim 7.5\%$ of the total depth 
for $p=3/2$, while in the case of $p=1$, the corresponding percentage 
is $40\%$.

One should note that our choice of $\tau_0$ and $T_e$ above
was rather arbitrary. However, the values of these parameters can be
meaningfully constrained from fits of the corresponding spectra of these
sources. In order to display in detail the constraints imposed by the combined
spectral - temporal analysis of the BHC data we now present fits of both
the spectra and the corresponding lags associated with the observations
of Cyg X-1 by Cui et al. (1997a,b).  As an example, we use here their 
results from observation \#6, when the source was in its high (soft) 
state while the other two reported set of data in the above references 
were from observations \# 3 and \# 15, during which the source was in a
transition between its soft and hard states.

In Figure 2 we have plotted the PCA (circles) and HEXTE (dots) data
covering an energy range $2-200$ keV from RXTE observations of Cyg X-1
in its 1996 high state (Cui et al. 1997a). On the same figure, 
in order to provide the tightest possible constraints to our models,
we have also plotted the BATSE data obtained by 
Ling et al. (1997) two years earlier while the source was in a similar 
high (soft) state, referred to by Ling et al. (1997) as the $\gamma_0$ 
state. By definition, in the high state, the soft ($\lsim 10$ keV) 
X-ray flux is high relative to the hard ($\gsim 30$ keV) flux. Such a 
distinct anti-correlation between the soft and hard bands was confirmed 
by the simultaneous monitoring of Cyg X-1 with the ASM/RXTE and BATSE 
(Cui et al. 1997a, Zhang et al. 1996). Because the BATSE observations 
of Ling et al. (1997) were indeed obtained during such a high state 
(note the almost identical fluxes between HEXTE and BATSE in the 
overlap energy range - the small difference could be attributed to 
uncertainties in the cross calibration between HEXTE and BATSE), 
we speculate that the state of the source was not very
different in these two observations, even though no simultaneous 
coverage of its soft energy band was at the time available (in 
any case these observations are served to {\sl constrain} rather 
than facilitate the fits of our models). 

In Figure 3 we have plotted the time lags between the energy bands 
$13 - 60$ keV and $2 - 6.5$ keV corresponding to the above data set
(i.e. of observation \#6) as a function of the Fourier period, with 
the data points re-grouped into logarithmically uniform bins in Fourier 
period. A linear dependence of the lag on the Fourier period in the
range $P \sim 0.03 - 3$ second is evident as in the GINGA data of 
Miyamoto et al. (1988).

In both Figures 2 and 3 we present, in addition to the data, model 
fits to the spectra (in Figure 2)  and the associated lags (in Figure 
3) corresponding to configurations of different values of the parameter
$p$. Thus the solid lines, in both figures, correspond to coronae
with $p=1$, the dashed lines to $p=3/2$, while the dotted ones to 
a uniform distribution ($p=0$). The temperature of the plasma was 
taken to be $kT_e = 100$ keV. The rest of the parameters of the 
configurations are as follows: i.) For $p=1$, $n_i = n_1 = 4.35\times
10^{16}$ cm$^{-3}$, $r_1 = 10^{-4}$ light seconds,  with a Thomson 
depth $\tau_i = 0.2$. The total optical depth of the corona is 
$\tau_0 =1$.  ii.) For $p=3/2$ and $n_1 = 1.6 \times 10^{17}$ 
cm$^{-3}$. Its uniform inner core has the same radius as the $p=1$ 
case, but a Thomson optical depth $\tau_i = 0.07$. 
iii.) For $p=0$, $n = 10^{16}$ cm$^{-3}$, and $\tau_0 = 0.5$ leading
to $r_2 \simeq 7.5 \times 10^7$ cm. 

It is seen that over the energy range from 20 to 200 keV, the spectra 
corresponding to $p = 1$ and 3/2 are almost identical to those of the 
$p=0$ configuration (dotted curve). In fact, their $\chi^2$ values are 
7.8 and 8.1 respectively (with 8 degrees of freedom), comparable to 
that of the uniform source. In other words, these three models provide
equally good spectral fits to the data, an argument made from a purely
theoretical point of view also in KHT. One should note, however, that 
the total Thomson depths of the corresponding configurations are indeed
different in these three cases. On the other hand, the differences in 
the lags corresponding to these three configurations are too apparent
to require statistical analysis. The shapes of the curves are similar to
those in figure 1, but the magnitudes of the lags are slightly smaller,
because of the smaller gap between the reference energy bands
in this case. Obviously, the corona with $p=1$ is again favored by the
observation.

Similar lag analyses have been performed for a number of BHC sources
in a small number of observing epochs (compared to those of spectral
analyses of the same sources). Thus, analysis of the light curves 
of the source GX 339-4 obtained by Ginga (Miyamoto et al. 1991) and 
the source GRS 1758-258 obtained by RXTE  (Smith et al. 1997, Figure 4) 
have indicated a similar linear dependence of the lags on the Fourier 
period. In the latter case, the time lags of hard X-rays ($6 - 28$ 
keV) with respect to soft ones ($2 -6$ keV) extend to $\delta t 
\simeq 1$ sec for $\nu \sim 0.02$ Hz! In Figure 5, we present 
the lags associated with the high energy transient source GRO J0422+32,
which was observed by CGRO/OSSE  during outburst. Analysis of the 
associated lags between the hard ($75 - 175$ keV) and soft X-rays
($35 -60$ keV) in this data set, indicated a similar behavior 
(Grove et al. 1997, 1998) over the Fourier period range $P \sim 
0.1 - 100$ seconds. The time lags level-off at $\omega \sim 0.025$ 
Hz after reaching a value of $\sim 0.3$ seconds.
In Figure 5 we provide a tentative fit to
these data obtained from a calculation based on Comptonization in 
a model corona with $p=1, kT_e =100$ keV, $\tau_0 = 0.25, \, n_1 
= 2.5\times 10^{15}$ cm$^{-3}$, $r_1= 4.8\times 10^{-4}$ and $r_2 
= 6.4$ light seconds. Although these parameters can be determined
unequivocally only by fitting the simultaneous spectral and temporal 
data, the discussion in \S 3 will show that at least the parameters 
$p$ and $r_2$ can be estimated with the lag observation alone. 
Particularly, $r_2$ of the model corona can be estimated from the 
equation
\begin{equation}
r_2 \sim P_c/2\pi,
\end{equation}
where $P_c$ is the period at which the lag curve levels off, which is
$\simeq 40$ seconds in this particular case. These results suggest 
that the frequency - dependent hard X-ray time lags may be a common 
property of these sources, at  least during particular states of 
their emission.

While the data presented so far seem to favor Comptonization in 
coronae with $p=1$, there are indications of a density profile with 
$p=3/2$ in the recent Cyg X-1 data in its low (hard) state (Wilms 
et al. 1997) and during the transition states (Cui et al. 1997b).
In Figure 6 we display the time lags of hard  ($14.09 - 100$ keV)
with respect to soft ($0 - 3.86$ keV) X-rays based on the RXTE data
reported by Wilms et al. (1997). We also plot the time lags (solid
curve) resulting from a calculation of Comptonization in a model
corona with $p = 3/2, kT_e = 100$ keV, $\tau_0 = 1.5$, $n_1 = 2.53
\times 10^{16}$ cm$^{-3}$, $r_1 = 0.001$ and $r_2 = 5$ light seconds. 
While we need the simultaneous spectral data in order to determine 
the values of $T_e$ and $\tau_0$, it is clear that the density 
profile with $p = 3/2$ does provide a valid fit to the lag data, 
independent of the particular spectral fit.

More recently, Crary et al. (1998) reported the hard X-ray
lags of Cyg X-1 observed by CRGO/BATSE during a period of approximately
2000 days. The time lags are between the energy bands $50 - 100$ keV
and $20 - 50$ keV over the Fourier period ranging from $P \gsim 5$
seconds to as large as $\sim 130$ seconds. Again, the lags are 
frequency dependent, obeying the relation $\delta t \propto 
\omega^{-0.8}$. As we have seen in the Figures 1, 3 and 6, both 
density profiles $p = 1$ and 3/2 will lead to lags of the form 
$\delta t \propto \omega^{-q}$ with $q \simeq 1$ and 0.6 respectively. 
Since the result of Crary et al. (1998) is an average over a very 
long time span, it is not unreasonable to speculate that during this 
time the density profile of the corona in Cyg X-1 evolved between 
the configurations with $p=1$ and 3/2. This data set 
has extended the lag determination to the longest Fourier period  to 
date and discovered X-ray lags rising almost out to this Fourier period. 
According to Equation (3) given below, this result, if confirmed,  
implies an exceptionally large outer edge for the Comptonizing 
corona ($\sim 20$ light seconds!!).

\section {The Computation of Lags}

As indicated in the explicit examples and fits to the data given in 
the previous section, the information associated with the lags is 
largely independent of that of the spectra and necessary for providing
a complete description of the structure of Comptonizing sources, an 
issue to which spectral fits can contribute very little. The power
spectra can also provide information about the structure of the sources,
however, because their form can be affected by the modulation of 
the accretion rate (see end of this section), this information is not 
unequivocal. 

The lags associated with the light curves in two different energy 
bands are essentially the {\sl average} of the difference of the paths 
lengths (divided by $c$) followed by photons in the two 
different energy bands, the photons of larger energy having spent
longer in the hot plasma. For a uniform source, this difference is 
(roughly) some fraction of the scattering time of the photons in 
the medium (Payne 1980), and for media of small Thomson depth, 
$\tau_0 \sim 1$, it is of order of a fraction of the light crossing 
time of the scattering region. Since the size $R$ of this region
corresponds to a particular Fourier period $P \sim R/c$, in 
principle, lags of all periods $P$ are present in sources with any
type of spatial density distribution. What distinguishes the form 
of lags in sources with different spatial density profiles is the 
additional probability of scattering within a region of a given 
size $R$, $\cal{P}$($R$). Thus, for a uniform density configuration, 
$\cal{P}$($R$) $\propto R$ (for $R < r_2; \; r_2$ is the outer edge 
of the corona), leading to lags of the form  $\delta t \propto R \cdot 
\cal{P}$($R) \propto R^2 \propto P^2$ for $P \lsim r_2/c$; the 
lags become constant for larger periods, since no further scattering 
takes place at longer time scales (larger radii). For $p=1$, $\cal{P}$($R) 
\simeq$ constant, leading to lags  $\delta t \propto P$ as indicated 
by our simulations, while for $p=3/2$, $\cal{P}$($R$) $\propto R^{-1/2}$, 
leading to $\delta t \propto P^{1/2}$, again in agreement with the 
results of our Monte Carlo simulations presented in the figures above.

One should note that a conclusion following from the above arguments is
that the {\sl form} 
of the lags as a function of Fourier period, under the conditions 
discussed above, i.e. constant electron temperature and 
injection of the soft photons near the center of the corona, depends 
mainly on the probability of photon scattering within a given range of 
radii; therefore the lag dependence on the Fourier period $P$, is 
characteristic of the electron density profile alone and, in our 
opinion, it could serve as a probe of it. The reader should note 
the distinct difference in information content 
between the lags and the spectrum: the latter 
depends simply on the total probability of scattering through the 
entire corona, while the former on the differential probability of
scattering within a given range of radii. 
Because the lags are simply a manifestation of the difference in 
average path of photons in two different energies, they also depend
on the position of the photon source, as it will be exhibited through a 
number of examples in the next section.

The dependence of the lags on the Fourier period associated with the 
plasma configurations discussed in the previous section and depicted
in the appropriate figures can be derived more formally using the 
Green's function of the associated problem, i.e. the form of the high 
energy light curves in response to an instantaneous input of soft 
photons at their center. In KHT, it was shown by numerical 
simulation that the Green's function of this problem for a configuration 
with $p=1$ has a power law shape, with index close to $-1$ which
becomes flatter with increasing total Thomson depth and photon energy. In
both cases there is a cutoff corresponding to the time scale, $\beta$,
characteristic of the photon escape time from the system. These
light curves can be approximated fairly well analytically by the 
Gamma function distribution
\begin{equation}
g(t) = \cases{t^{\alpha -1}e^{-t/\beta}, &if $t\ge 0$; \cr
                0, &otherwise, \cr }
\label{gamma}
\end{equation}
where $t$ is time; $\alpha > 0$ and $\beta > 0$ are parameters determining
the shape of these light curves.

The light curves from a uniform electron corona corresponds to $\alpha 
\simeq 1$ so that the Gamma distribution function reduces to an 
exponential. Because the hot electron configuration in this case is 
assumed to be confined to the vicinity of the compact object, the 
corresponding value for $\beta$ is assumed to be of order $10^{-3}$ 
sec, though more extended uniform configurations can in principle be 
analyzed by letting the value of this parameter be proportionally 
larger. A corona with density profile corresponding to $p > 0$ is 
inhomogeneous and the properties related to this specific characteristic 
manifest for large values of the  ratio of its inner and outer radii 
$r_2/r_1$. Motivated by the observations we require $r_2 \sim 1$ light 
second, which corresponds to $\beta \sim 1$ second, and with a value 
for $\alpha$ small compared to 1. The Fourier transform of the 
function $g(t)$ is
\begin{equation}
G(\omega) = \displaystyle{{\Gamma(\alpha) \beta^{\alpha}}\over
              {\sqrt{2\pi}}}(1+\beta^2\omega^2)^{-\alpha/2}
              e^{i\alpha\theta}.
\label{fourier}
\end{equation}
where $\Gamma(x)$ is the Gamma function and $\omega = 2\pi\nu = 2\pi/P$ is
the Fourier circular frequency;  $\theta$ is the phase angle that we 
are interested in and is given by
\begin{equation}
\tan{\theta} = \beta\omega.
\end{equation}

If we consider two light curves in two energy bands distinguished by
different values of $\alpha$ and $\beta$, say $\alpha_1$, $\beta_1$ and
$\alpha_2$, $\beta_2$, the time lag between them will be given by their
phase lag $\theta$ divided by the corresponding frequency $\omega$,
i.e.
\begin{equation}
\delta t = \displaystyle{1\over \omega}~[\alpha_2\arctan(\beta_2\omega) -
\alpha_1\arctan(\beta_1\omega)]
\end{equation}
For large periods $P$ (small $\omega$) or $\beta_1 \omega,~\beta_2
\omega \ll 1$, $\delta t$
approaches the constant $\alpha_2\beta_2 - \alpha_1\beta_1$. On the 
other hand, for small $P$, or $\beta \omega \gg 1$, $\delta t \simeq 
(\alpha_2-\alpha_1)P/4$. In the case of a uniform corona for which 
$\alpha_2 = \alpha_1 = \alpha \simeq 1$, expanding Equation (7) 
to higher order in $\omega$,
we obtain $\delta t \propto \alpha(1/\beta_1 -1/\beta_2)P^2$. The 
transition from latter to the former occurs at
\begin{equation}
\beta \omega \sim 1 ~~~~{\rm or}~~~~ P_c \sim 2 \pi \beta.
\end{equation}
This immediately gives Equation (3) if we remember that the outer
radius $r_2$ is of the same order as $\beta$.

The analytic results presented above are given in graphic form in figures
7a and 7b. In Figure 7a, we plot two pairs of light curves associated 
with the function $g(t)$, one with $\alpha_1 = 0.1$,
$\alpha_2 = 0.2$, $\beta_1 = 1$ and $\beta_2 =1.25$, the other with
$\alpha_1 = \alpha_2 = 1.0$, $\beta_1 = 0.001$ and $\beta_2 =0.0025$. 
The former represents the light curves resulting from a corona with 
$p=1$ density profile and $r_2 \simeq 1$ light second, while the latter 
from a uniform one with $r_2 \simeq 10^{-3}$ light second. The time lags
between these two pairs of light curves are computed as indicated above
and presented in Figure 7b. The dependence of the lag shape to the 
parameters $\alpha$ and $\beta$ is apparent in these figures: For 
pure exponential light curves $\alpha = 1$ (or more precisely 
for light curves with $\alpha_1 = \alpha_2$), the time lag is 
proportional to $P^2$ for $P < P_c \simeq 0.006$ second, and 
turns to a constant value for larger $P$, while the value of the lags 
at this point is of order $\beta_2 - \beta_1 \simeq 1$ millisecond. 
For the curves with $0 < \alpha < 1$, the time lag is proportional 
to $P$ for $P< P_c \simeq 6$ second 
and turns to a constant for large $P$. In both cases, the level-off
point is $P_c \simeq 2\pi\beta$ with the value of $\beta$ appropriate
for each case, however, the time lag resulting from 
the purely exponential light curves has no portion linear in $P$. 
Thus the existence of a linear portion in the time lag curves 
obtained from fits to observations clearly favors light curves which
include power law sections as those of Eq.(\ref{gamma}).
As discussed in KHT, such a form for the response functions 
is a signature of a non-uniform density distribution of the 
Comptonizing corona and the values of $\alpha$ and $\beta$ are 
closely related to the physical size and density distribution of 
the source corona.

Having obtained the Fourier transform $G(\omega)$ (Equation 
\ref{fourier}) of the light curve $g(t)$ (Equation \ref{gamma}), 
one can easily compute the Fourier transform associated with a light
curve consisting of the sum of two such light curves separated
by an interval $t_0$, i.e. the light curve $g_1(t) = g(t) + g(t-t_0)$.  
According to the well known Fourier analysis theorems, this will
consist of the sum $G(\omega)+G_0(\omega)$, with $G_0(\omega)
= G(\omega)\, e^{i \omega t_0}$. Therefore, the presence of 
the additional shot induces only a change in the {\sl phase} of 
the corresponding  Fourier transform, the latter becoming $\theta 
+ \omega t_0$. Since the change in phase associated with the second
light curve, $g(t-t_0)$, is the same for photons which belong in 
two different energy bands (the light curve of each band is shifted 
by $t_0$), their {\sl phase difference}, and hence their time 
lags, are unaffected by the presence of additional shots. Considering
that any light curve $f(t)$ can be written as the  convolution of 
$g(t)$ with a modulating function $M(t)$, by the arguments given 
above, such modulation will not affect the {\sl phase differences} 
and hence the time lags between two different
energy bands. However, it will have a great impact on the resulting
power spectra which now are proportional to $\vert M(\omega)\cdot 
G(\omega) \vert ^2$, where $M(\omega)$ is the Fourier transform of
$M(t)$. 

It becomes apparent, therefore, that the additive nature of the phases
associated with the Comptonization process makes phase differences,
and hence the corresponding time lags, immune to any particular modulation 
of the rate at which matter is accreted onto the compact object.
As such they map out properties which are inherent to the Comptonizing
hot electron plasma rather than to the modulation of the accretion
rate (usually invoked to account for the PSD), 
in as much as the latter does not change significantly the plasma's 
overall properties. In this respect, Miller (1995) has argued that
oscillations of the scattering corona do preserve the phase relations
established by the Comptonization process, provided that their period
of oscillation is longer than the scattering time responsible for the
formation of the lags. Our point of view in the present note is  
slightly different: in our simple models we posit that the observed lags 
are in fact produced by scattering in the corona and that their magnitude 
provides an estimate of its size. The sizes thus deduced by our models
are much larger ($\lsim 10^{11}$ cm) than the size of the horizon of the
putative black hole, where presumably most of the accretion energy 
is converted in radiation; this discrepancy between the estimated 
(through the lag measurements) and the expected (on energetic arguments) 
sizes of the X-ray emitting region is precisely the controversial aspect
of the observed lags. In an attempt to ameliorate this discrepancy, 
Nowak et al. (1997) have proposed models of waves propagating inward 
in the accretion disks near the compact object, which at the same time
are associated with radiation emission; 
they further postulated that the emitted radiation becomes hotter at 
smaller radii, thus giving rise to lags between soft and hard X-rays. 
As the authors themselves point out, the problem with this type of 
model is that the wave speeds needed to produce lags of the 
observed magnitude are $c_s \simeq 0.01 c$, i.e. much slower than 
those typical of the dynamics or radiation transfer in this region;
it is also not obvious whether such waves, if of random phases, would  
yield the high values of the coherence function observed in this 
class of sources (see next paragraph). Therefore, in either way, 
the size of the observed lags indicates that some of our cherished 
notions concerning the structure and dynamics of these objects are 
not correct in a big way!

The properties of the Comptonizing coronae are generally expected to 
be time variable
and their variability should have some effects on the corresponding
phases. The issue of the preservation of phases in time varying 
coronae is, to a large degree, related to the observed high values
of the coherence function (CF) of the light curves of accretion
powered high energy sources (see Vaughan \& Nowak 1997; HKT), which 
effectively is the normalized cross spectrum at two different energy
bands, averaged over the duration of the observation. As argued in 
HKT, the linearity of the Comptonization process, leads naturally 
to high values ($\simeq 1$) for the CF in plasmas with parameters
which do not change in the duration of observation. HKT have explored 
the dependence of CF to variations in the plasma temperature $T_e$;
as shown in their figure 2, changes in $T_e$ by a factor of 2 over 
the duration of observation lead to a uniform decrease in the value 
of the Coherence Function ($CF \simeq 0.8$ as opposed to $CF \simeq 
1$ for a constant $T_e$) over the entire 0.1 - 100 Hz frequency range.
This uniformity in the CF decrement is 
indicative of the fact that the relative sign of the phases is rather
well preserved despite the change in the electron temperature (this 
last statement depends on the energy bands used in the computing the 
coherence function; in HKT they were both below the minimum 
temperature attained by the electrons in the course of the variation
of their temperature). Observations indicating $CF \simeq 1$ 
(Vaughan \& Nowak 1997; Nowak et al. 1997) strongly suggest 
that the temperature of the scattering plasma has remained constant 
over the entire observation interval.

Thus, under the assumptions of the present calculations (i.e. 
Comptonization as the main process of high energy emission, uniform 
temperature, non-uniform density, central soft photon injection), 
the time dependence of the photon flux, i.e. the light curves, at various
energies can be used to determine the size of the hot electron corona
and  map its radial density distribution (i.e. determine the index $p$).
This fact provides the possibility of deconvolution of the density 
structure of these coronae through a combined spectral-timing analysis
similar to that indicated above.

\section{Lags and The Soft Photon Source Location}

In the previous section we discussed the general notions associated
with lags induced by Comptonization in inhomogeneous media of spherical 
symmetry and their dependence on the Fourier period, which was 
derived both heuristically and formally for specific cases. It was 
also argued that the form of the lags so derived and exhibited in 
Figures 1, 3, 4, 5 and 6 depends, in addition, on the assumption that 
the injection of the soft photons takes place near the center of 
the hot, spherical electron configuration. It is easy to see that 
if the latter assumption is not true the resulting lags could have  
different form than those computed above: if the source is of finite 
extent or not located at the center, the scaling concerning 
the probability of photon scattering at a given range of radii 
would be different from that given above, leading to different 
scaling of the lags as a function of Fourier period. It is hence
conceivable, under these conditions (extended or non-central soft
photon source) and if the coronae are optically thin, that one could
receive soft photons emitted from the farther edge of the 
soft photon source at a later time than photons originating in its near 
side, which did increase their energy by scattering on the way 
to the observer, thereby introducing negative lags in the system.
The effects of a centrally located source of finite extent $r_{in}$ 
were discussed in HKT, where it was pointed out that it would lead
to loss of coherence on time scales $r_{in}/c$. This loss of coherence
is generally manifest with the presence of negative lags in the 
appropriate Fourier frequency range. 

Therefore, the form of the observed lags conveys information not only
about the density distribution of the hot electrons but also about
the distribution of the source of soft photons. The effects of the
finite distribution of the soft photon source manifest in the loss 
of coherence
of the scattering medium over the periods appropriate to the light
crossing time of the soft photon source, as discussed in HKT.
In this section we present the lags corresponding to a number of 
arrangements different of those examined heretofore, in order to 
substantiate the above arguments and in order to exhibit the 
form of the lags in certain geometries which are favored by 
popular models of these sources. 

We hence consider the location of the soft photon source to be 
{\sl external} to the hot, spherically
symmetric cloud and arranged in a ring geometry of finite width 
and inner radius $r_{in} = 4$ light seconds, in order to simulate 
injection of soft photons by an outlying
cool accretion disk. The hot electron cloud responsible for the 
Comptonization is assumed to be spherically symmetric, centered at 
$r=0$, with electron temperature $T_e =100$ keV, outer radius $r_2 = 
1$ light second and total Thomson depth $\tau_0 =1$. We further consider 
two different cases for the density distribution of the hot electrons: 
i.) A corona with $p=0$ (uniform), of density $n = 1.5 \times 10^{13}$ 
cm$^{-3}$ and ii.) A corona with a $p=1$ density profile, of inner radius 
$r_1 = 1.25 \times 10^{-4}$ light seconds and density $n_1 = 4 \times 
10^{16}$ cm$^{-3}$. We also consider two different cases for the 
outer radius of the ring, $r_{out} = 4.001$ and 4.5 light seconds 
respectively, in order to study the effects of the finite size of
soft photons on the structure of the lags. The soft photons 
are drawn from a black body distribution of temperature $T_r = 1$ 
keV, while the lags are computed between the following energy bands:
$1 - 5$ keV (band 1), $5 - 10$ keV (band 2), and $10 - 50$ keV
(band 3).  

In figures 8, 9, 10 and 11 we present the results of our simulations.
Figures with the subscript $a$ exhibit the form of the light curves
for the different energy bands indicated above, while figures of 
subscript $b$ show the corresponding lags between the marked pairs 
of energy bands. In figures 8a, 8b we present the results for a 
corona with $p=1$ and a soft photon source of $r_{out} = 4.001$ 
light seconds; In figures 9a, 9b the results of a uniform corona ($p=0$) 
and soft photon source with $r_{out} = 4.001$ light seconds. In 
figures 10a, 10b the results for a corona with $p=1$ and soft 
photon source with $r_{out} = 4.5$ light seconds, while in figures 
11a, 11b the results for a uniform corona and a soft photon  source 
with $r_{out}  = 4.5$ light seconds. 

The consistency of our results has been tested by reproducing with 
high accuracy the analytic results of the previous section for
the case of central soft photon injection both for a uniform and 
a $p=1$ corona. The light curves were derived by employing the 
Monte Carlo code of Hua (1997) with $10^7$ photons per run and
collecting the photons emerging from the hot electron cloud
in all directions; the lags were subsequently computed from these 
light curves as discussed in the previous section. 

The light curves of figures 8a and 9a exhibit a sharp peak at 
times near $t=2$ seconds for bands 1 and 2. These are due to 
photons which traversed the entire length of the corona through 
its center exiting on the opposite side with only a small gain 
in energy. The photons in band 3, having suffered a larger number 
of scatterings, have apparently followed very different paths and 
hence do not exhibit this feature. The corresponding lags are given
in figures 8b and 9b, marked by the energy bands involved; they
are of apparently different form than those obtained from central 
injection of soft photons. Their magnitude is also smaller than 
that obtained in our figures $1-6$ of the previous section, with 
the exception of the lags between bands 3 and 1 in the lowest 
frequencies ($\lsim 1$ Hz), which are indeed commensurate with 
the light crossing time through the corona. Of interest
is also the presence of some {\sl negative} lags (filled circles)
at certain frequency ranges, possibly related to some backscattered
photons. The general dependence of lags on 
Fourier frequency in these two cases is also different from that 
associated with central soft photon injection, though the $2 - 1$ 
curve in figure 8b does exhibit a general trend similar to the 
linear dependence of the $p=1$ coronae associated with central 
photon injection; on the other hand the form of the $3 - 1$ lag
curve is distinctly different from that corresponding to soft photon
injection.

Figures 10 and 11 exhibit the effects of the finite size of the 
soft photon source. In both the $p=1$ and the $p=0$ cases, the 
source's finite extent manifests  itself in the 
light curves as a plateau of duration 0.5 seconds starting again
at $t = 2 $ seconds, i.e. the light crossing time through the 
center of the corona. This feature is absent from the light
curves of band 3 for the reasons given above. The corresponding
lags are now substantially different in shape from those of the  
corresponding figures 8b and 9b, as indicated by the presence of 
negative lags at progressively low periods (especially in fig. 10b),
in qualitative agreement with the arguments made above.
These can be understood as the mixing of hard and soft photons
due to the finite size of the soft photon source which becomes 
more pronounced as the size of the source increases. It is
apparent that these lags are very different from those observed
in the sources discussed in the previous section, thus making
a strong case for the injection of the soft photons at radii
considerably smaller than the extent of the Comptonizing coronae.

\section{Conclusions}

The main results of the present investigation are the following:  

1)  Fourier frequency dependent hard X-ray lags appear to be  a common
property of accreting, galactic black hole candidates. 
Observed first in the hard spectral state of  Cyg X-1 (Miyamoto et al.
1988), they appear to be present in its soft  state too (Cui et al. 
1997b) and also in a number of other BHC sources. Of particular 
interest is their apparent increase with period, out to $P \simeq 
130$ sec in the case of Cyg X-1. Their observed dependence on the 
Fourier frequency $\nu$ has usually the form $\delta t \propto 
\nu^{-q}$, with $q$ varying between $\simeq 0.6 - 1.0$. 

2)  The simplest interpretation of the observed lags is that they are
due to the Comptonization process. Then, their magnitude provides an
estimate of the size of the scattering medium, a quantity inaccessible
to spectral modeling. Furthermore, their observed Fourier frequency 
dependence can be accounted for by imposing the additional assumption
that the Comptonizing plasma is inhomogeneous; the index $q$ of their 
Fourier frequency dependence is then found to be related in a simple way
to the spatial density profile of the scattering electrons, i.e. the 
index $p$ (Eq. \ref{density}). For $\nu \gsim c/r_2$ this relation,
as derived from our analytic and numerical calculations, appears to be 
of the form $q \simeq 2 - p$.

3)  Simple analytic models can be used to compute the  
lags resulting from Comptonization in spherically symmetric, inhomogeneous 
coronae as a function of the Fourier frequency. We have also used these 
results to show that these lags, in contrast to the PSD, are generally 
not affected by variations in the matter accretion rate onto the 
compact object; as a result they probe properties inherent to the 
scattering electron cloud rather than to fluctuations in the accretion 
rate. We have also indicated that our analytic and numerical results,
through the relation between the density index $p$ and the lag index 
$q$, can be used to {\sl map} the spatial density structure of this 
class of objects, providing a probe of the dynamics of accreting high
energy sources beyond that afforded by spectral analysis alone.
As such, the observed lags associated with BHC sources are in clear 
disagreement with our models of uniform coronae, while in certain cases 
are consistent with the density profiles usually attributed to ADAF
(Narayan \& Yi 1994). 
In most cases they appear consistent with density profiles of the 
form $n(r) \propto 1/r$, which is not associated with any obvious 
model of accretion dynamics. The existence of apparently more than 
one (simple) form for the observed lags raises the possibility of 
existence of distinct {\sl timing states} for these sources, similar
to their well known spectral ones. The (albeit) small number of 
lag analyses to date indicate that the timing and spectral states 
are not necessarily correlated, thus raising the possibility of an 
additional dimension in the study of the issue of the different
states of BHC sources.

4)  One of the most puzzling and controversial results of the present 
study is the large sizes of the Comptonizing coronae of BHC sources, 
implied through our models, by the magnitudes of the observed 
lags; the sizes derived are in excess of $10^{11}$ cm, several orders
of magnitude larger than the horizon size of the putative black holes,
where most of the accretion energy is thermalized. We do not have, as
yet, an obvious mechanism by which the accretion energy can be deposited
to energetic electrons over such distances. We consider an account
of the observed lags one 
of the major reasons for vigorously pursuing studies along these lines.
On the other hand, it maybe possible for more complicated models 
to produce the observed lags by processes taking place in a much 
smaller region, closer to the black hole horizon (e.g. Nowak et al. 
1997); however, given the state of these models, we believe
that Occam's razor favors, for the time being at least, 
the ones presented herein. 

5)  The position and size of the soft photon source have strong
effects on the form of the lags as a function 
of Fourier frequency; the results of our simulations indicate 
that an soft photon source exterior to the corona produces lags
which are, in general, of considerably different form than those 
due to a source located at its center. The differences become 
particularly pronounced if the soft photon source is of finite 
extent, a fact manifested by the presence of negative lags 
at increasingly larger Fourier periods and which, as argued in HKT, 
would greatly reduce their coherence 
function. The data from the limited number of sources presented in 
section 2 seem to favor central rather than exterior soft photon
injection.

The results we have presented above were produced under the 
assumption that the scattering electron cloud is spherically
symmetric and optically thin. One expects that at least the 
first of these assumptions would be violated in realistic sources.
However, as long as their overall geometry is not very different
from spherical, i.e. the ratio of their largest to their 
smallest dimensions is not much greater than one, the general
conclusions reached above would still hold. If the Thomson depth 
of these configurations becomes considerably larger than 1, it 
would first impact the resulting spectra and would also make the
parameter $\alpha$ in Equation (\ref{gamma}) larger, leading 
to flatter PSDs; such a correlation between the spectral and 
PSD indices should be searched in the existing data. Finally, if
the source is very asymmetric and sufficiently opaque, both the 
spectra and the lags should have a strong dependence on the 
observer's orientation; such a dependence could be uncovered by 
searching for correlation of the spectral and timing properties of 
these systems with the phase of 
the binary orbit. We are not at present aware that any such 
a dependence has been observed in any of the sources discussed
above. 

In conclusion, timing analysis of BHC sources 
through the study of time lags appears to 
be a powerful tool, complementary to the more commonly employed 
spectral analysis. Both are necessary, however, in order to determine
the properties of the Comptonizing coronae of accreting high energy
sources: the latter in order to provide the temperature and optical
depth of the coronae, while the former in order to provide estimates
of the physical size (assuming that the lags are due to Comptonization)
and spatial distribution of the hot electrons. These results 
supplemented by further 
timing analysis through the study of PSDs and moments of the 
observed light curves are expected to provide additional constraints
which will allow the unequivocal determination of the dynamics of 
accretion onto the compact object.

\section*{Acknowledgments}

The authors would like to thank J.C. Ling, J.E. Grove, S.N. Zhang, J.
Swank and W. Focke for useful discussions.

\newpage

\centerline{\bf FIGURE CAPTIONS}

{\bf Figure 1.} 
The time lags of hard X-rays ($15.8 - 24.4$ keV) with respect
to soft ones ($1.2 -5.7$ keV) as a function of Fourier period, obtained 
from the analysis of Ginga data of Cyg X-1 (Miyamoto et al. 1988). 
The circles and dots represent the measured positive and negative lags 
respectively. The dash-dotted curve represents the lag predicted based 
on analytical 
Comptonization model given in Miyamoto et al. (1988) and assumed corona
electron density $10^{16} {\rm cm}^{-3}$.  The three other curves are
calculation results for Comptonization in the model coronae that produce
the energy spectra shown in Figure 2.

{\bf Figure 2.} 
Three calculated energy spectra which fit equally well the
Cyg X-1 data (crosses) in its $\gamma_0$ state observed by CGRO/BATSE
in 1994 (Ling et al. 1997). These spectra result from Comptonization in
coronae with the same temperature but different optical depths
and density profiles: dotted - $p=0, \tau_0=0.5$, solid - $p=1,
\tau_0=1.0$ and dashed - $p=3/2, \tau_0=0.7$. The dotted and dashed
curves are slightly displaced to separate the otherwise nearly
identical curves. Also plotted are RXTE/PCA (circles) and HEXTE (dots)
data from the same source observed during its high state in 1996
(Cui et al. 1997a). 

{\bf Figure 3.} 
The time lags of hard X-rays ($13 - 60$ keV) with 
respect to soft ones ($2 -6.5$ keV) resulting from Comptonization 
in the same coronae that produce the energy spectra shown in Figure 
2. The dotted, solid and dashed curves, as in Figure 2, represent 
the density profiles $p=0, 1$ and 3/2 respectively. The time lag 
between the same energy bands based on RXTE data from Cyg X-1 
(Cui et al. 1997b) are also plotted in the figure.

{\bf Figure 4.} 
The time lags of hard X-rays ($6 - 28$ keV) with respect to soft ones
($2 -6$ keV) as a function of Fourier period for the source GRS
1758-258. This is converted from the phase lag vs frequency relation 
obtained by Smith et al. (1997). Dots are negative lags.

{\bf Figure 5.} 
The time lags of hard X-rays ($75 - 175$ keV) with respect to soft ones
($35 -60$ keV) as a function of Fourier period for the source
GRO~J0422+32, obtained by Grove et al. (1997) from OSSE observation. The
solid curve is obtained from a calculation based on Comptonization in
a model corona with $p=1, r_2 = 1.9 \times 10^{11} {\rm cm},
kT_e =100$ keV and $\tau_0 = 0.25$.

{\bf Figure 6.} 
The time lags of hard X-rays ($14.09 - 100$ keV) with respect to soft
ones ($0 -3.86$ keV) as a function of Fourier period for the source 
Cyg X-1 based on RXTE data (Wilms et al. 1997). The solid curve is a 
result from the calculation of Comptonization in a corona with $p = 3/2,
kT_e = 100$ keV, $\tau_0 = 1.5$ and $r_2 = 5$ light seconds.

{\bf Figure 7.}
(a) Two pairs of light curves corresponding to two sets of values of
$\alpha$ and $\beta$ in Equation (2). The pair with $p= 1$
represents the light curves from a cloud with $1/r$ density profile,
the other pair from a uniform one. The solid curves correspond to
the light curves in the lower energy bands. (b) The time lags based on
Equation (7) between the two curves in each pair in Figure 7. The dashed
curves are times lags obtained by numerical Fourier transformation 
with finite time resolution.

{\bf Figure 8.}
(a) The light curves corresponding to 3 energy bands (band 1: $1 - 5$ 
keV; band 2: $5 - 10$ keV; band 3 $10 - 50$ keV) as marked on the 
figure for a soft photon source exterior to the hot electron cloud.
The source is located on a ring of radii $r_{in} = 4.0, \; r_{out}
= 4.001$ light seconds. The hot electron corona is concentric to 
the ring with an electron 
temperature $T_e = 100$ keV, total Thomson depth $\tau_0 = 1$
and outer radius $r_2 = 1$ light second. It has a density profile 
with $p =1$,  $r_1 = 1.25 \times 10^{-4}$ light second and $n_1
= 4 \times 10^{16}$ cm$^{-3}$.  (b) the lags as a function of Fourier
frequency between bands 3,1 and 2,1 as marked on the figure for the 
configuration corresponding to (a).

{\bf Figure 9.}
(a) Same configuration and parameters as for fig. 8a except for a 
uniform corona ($p=0$) of $n = n_1 = 5 \times 10^{13}$ cm$^{-3}$.
(b) Same as in fig. 8b for the configuration corresponding to (a).

{\bf Figure 10.}
(a) Same configuration and parameters as for fig. 8a except for a
ring with radii $ r_{in} = 4.0, \; r_{out} = 4.5$ light seconds.
(b) Same as in fig. 8b for the configuration corresponding to (a).

{\bf Figure 11.}
(a) Same configuration and parameters as for fig. 9a except for a
ring with radii $ r_{in} = 4.0, \; r_{out} = 4.5$ light seconds.
(b) Same as in fig. 9b for the configuration corresponding to (a).

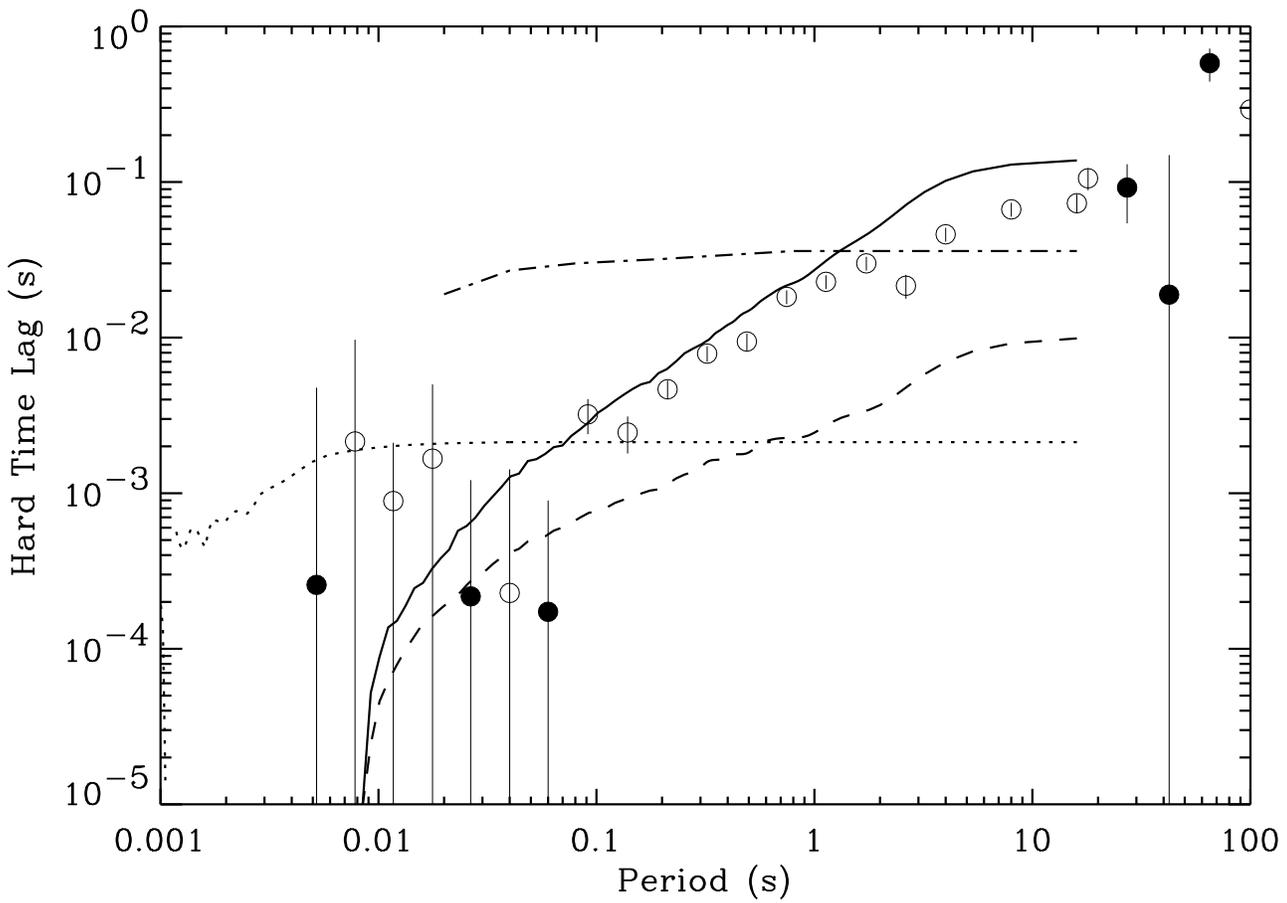
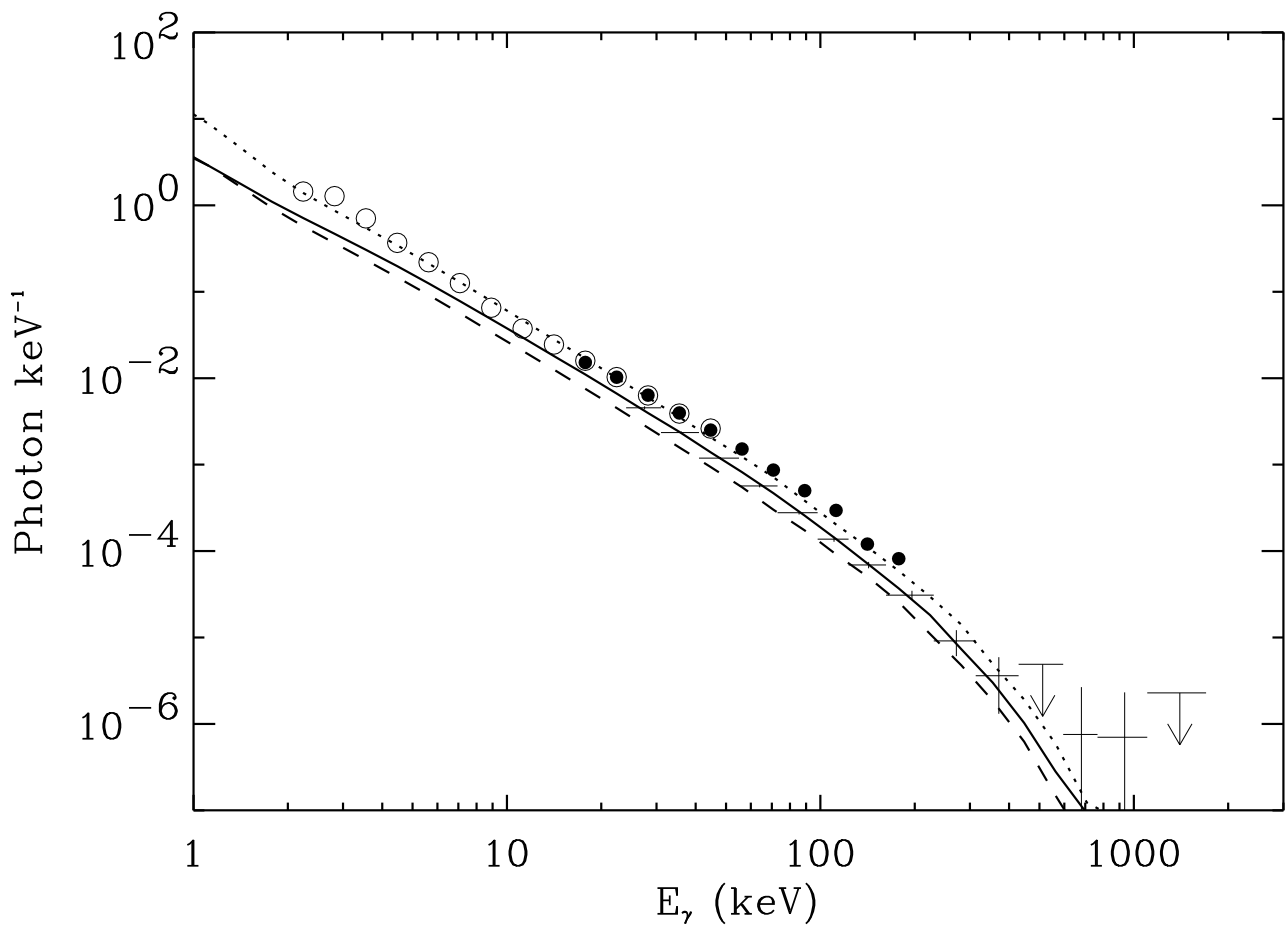
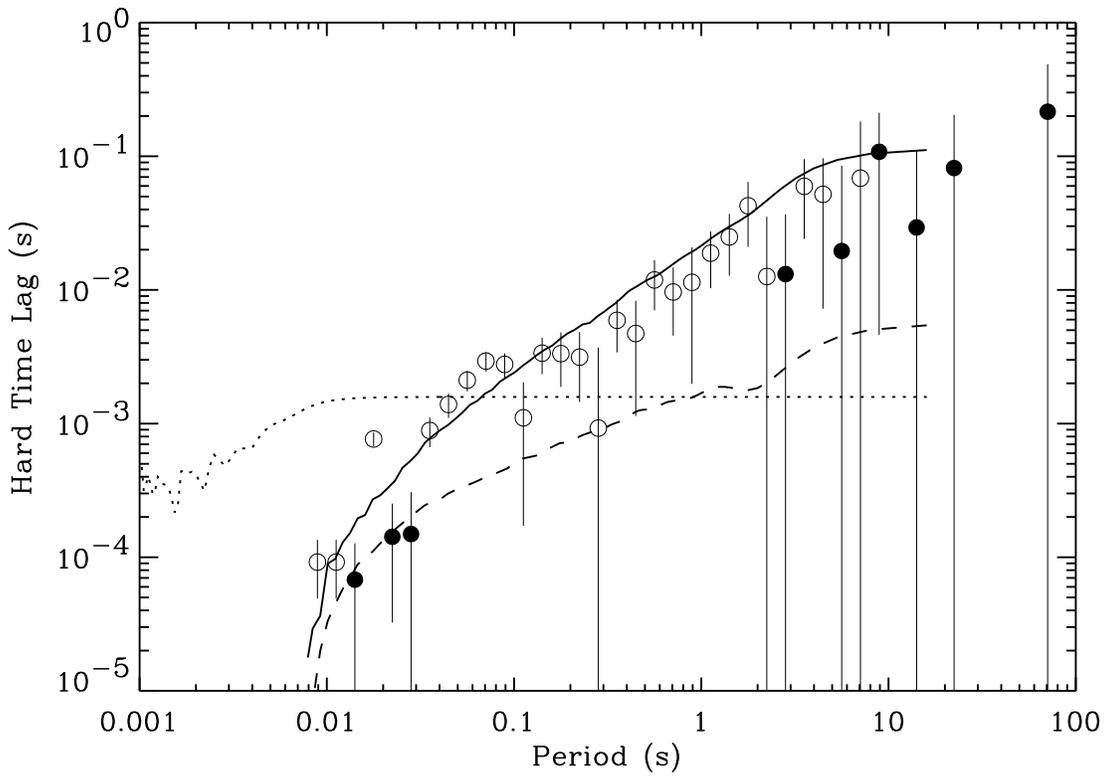
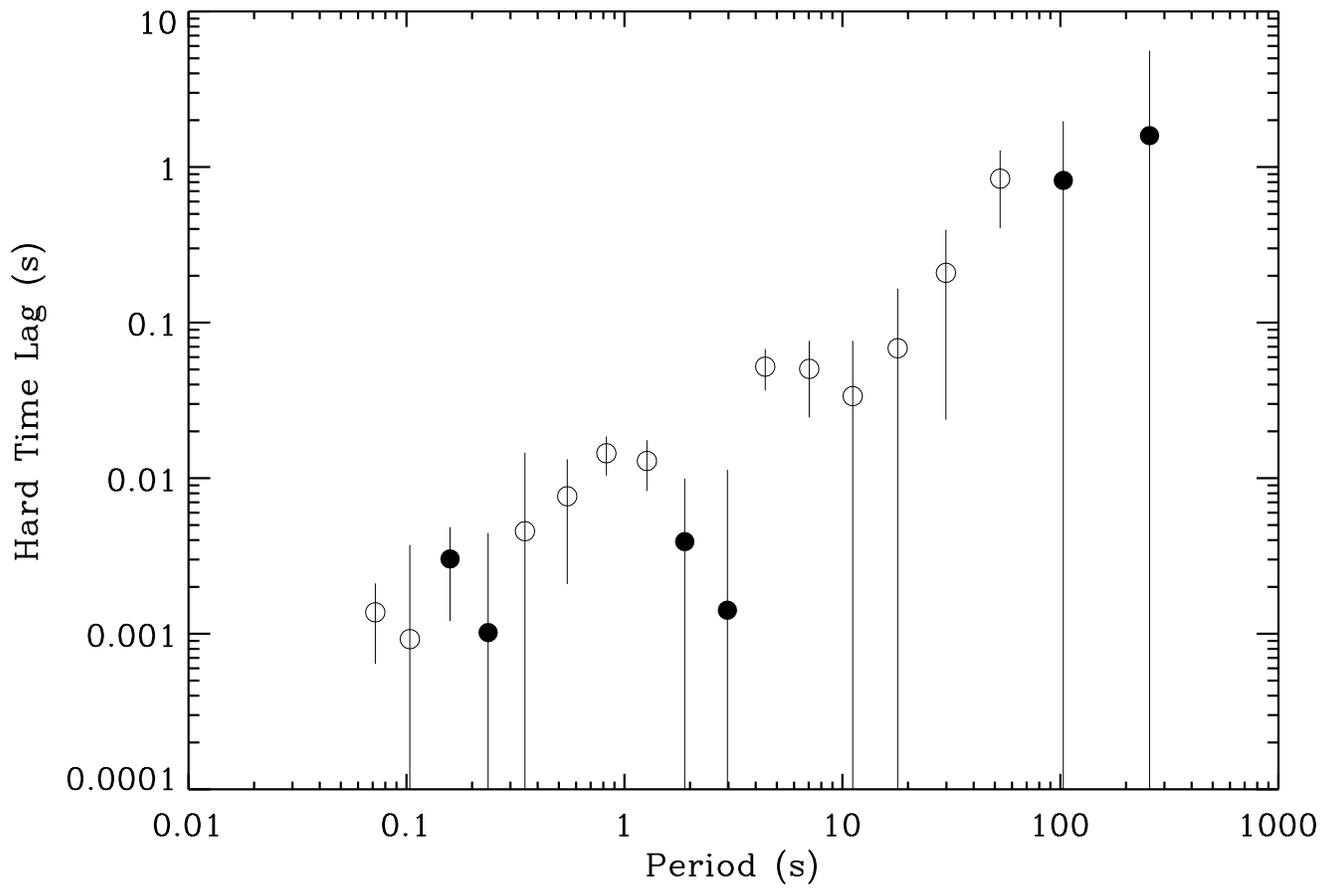
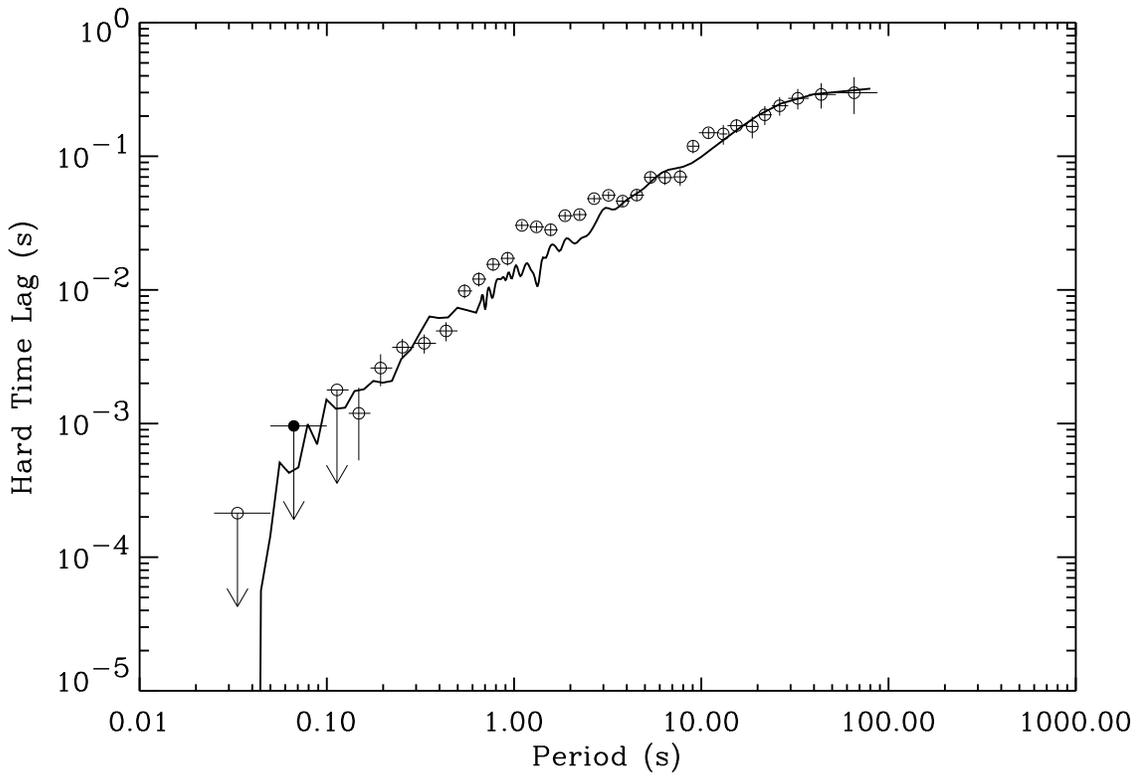
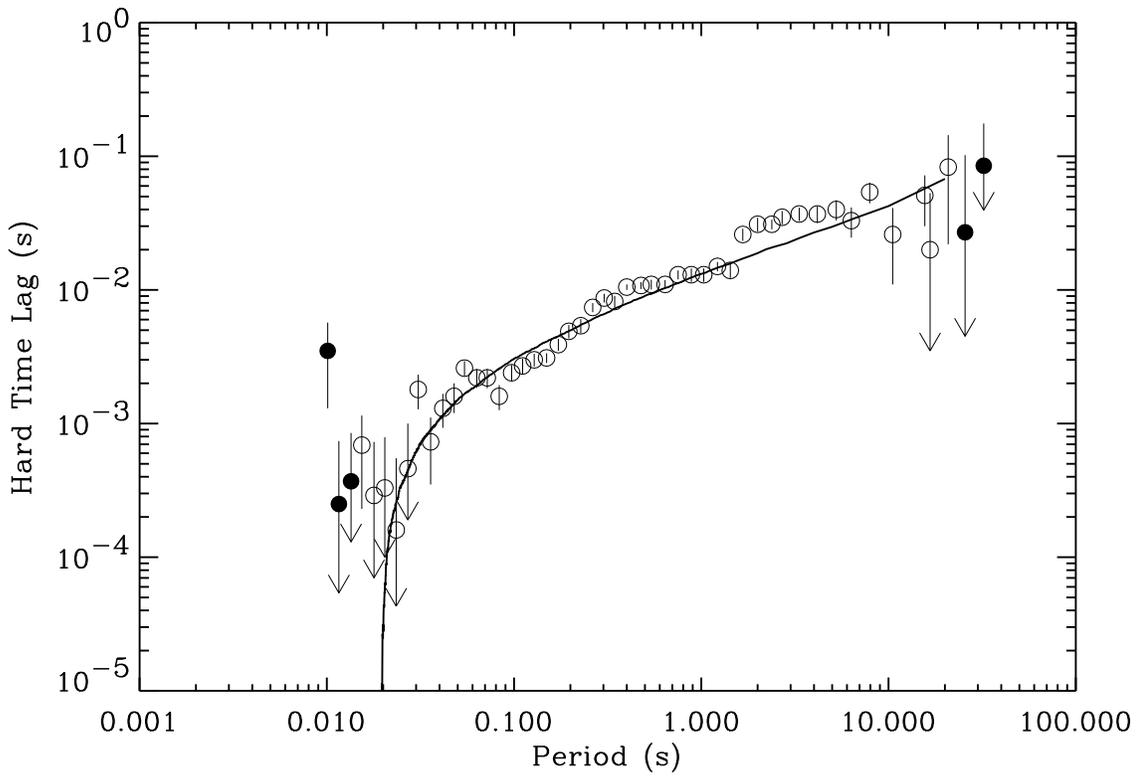
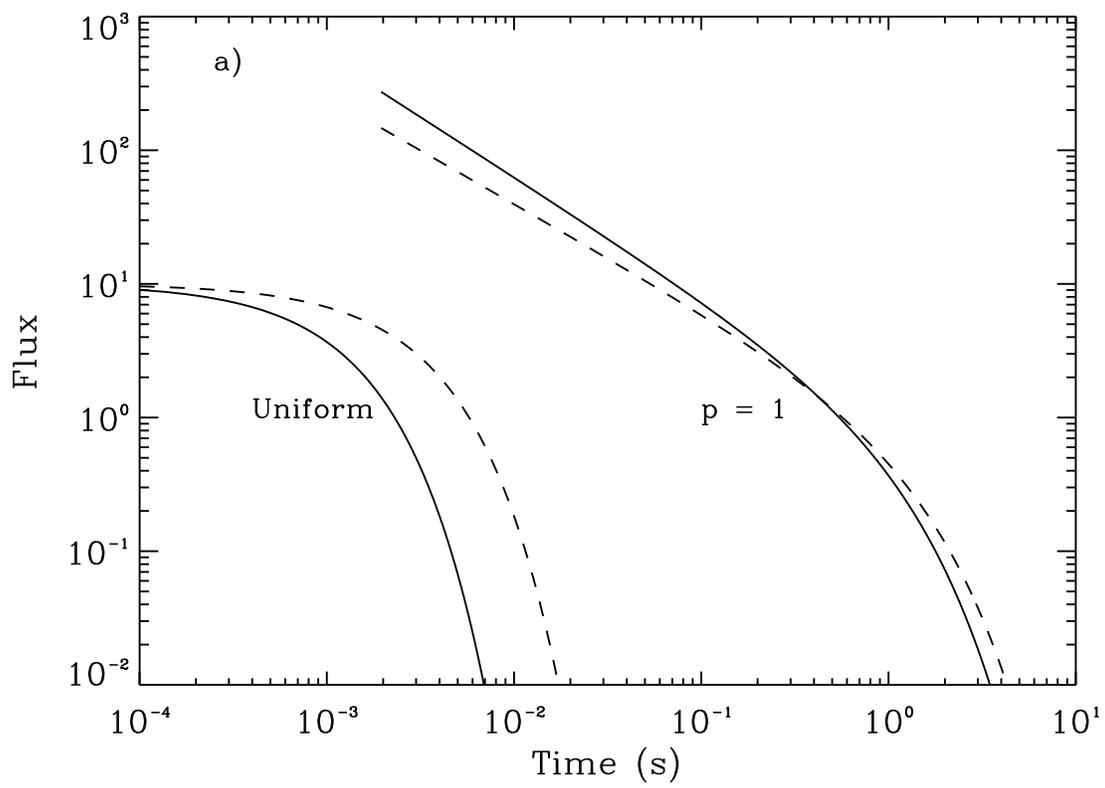
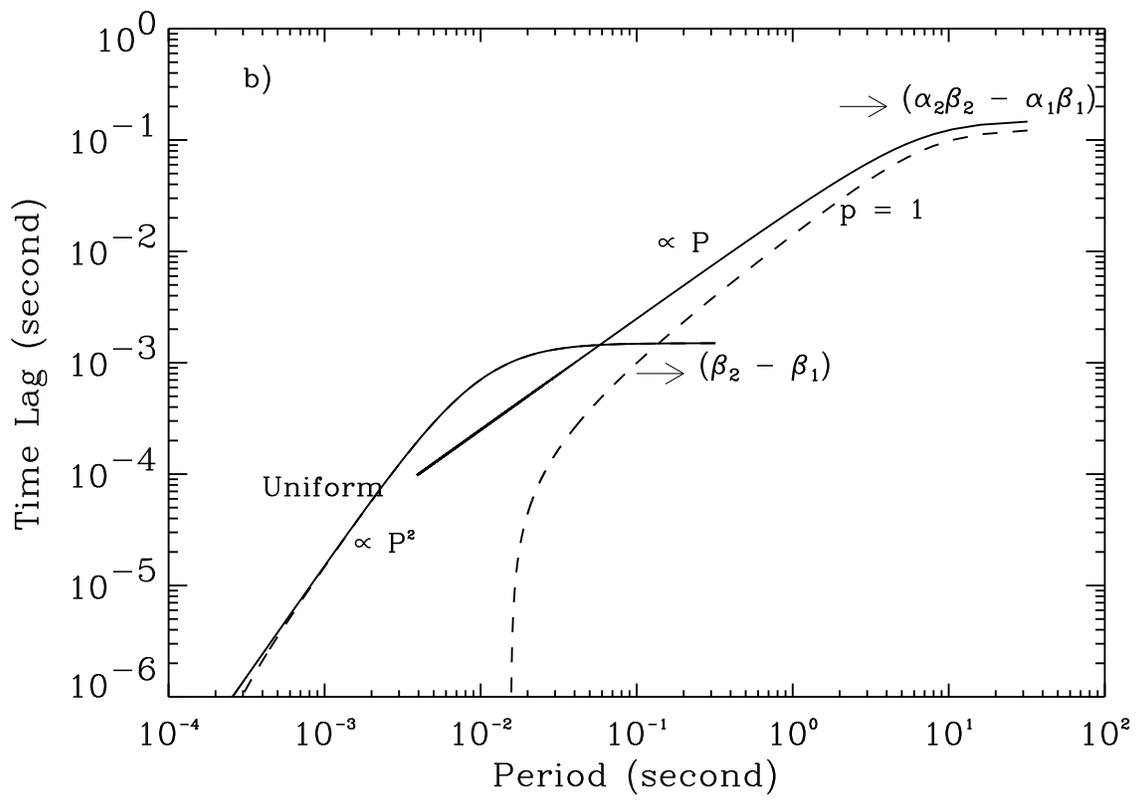
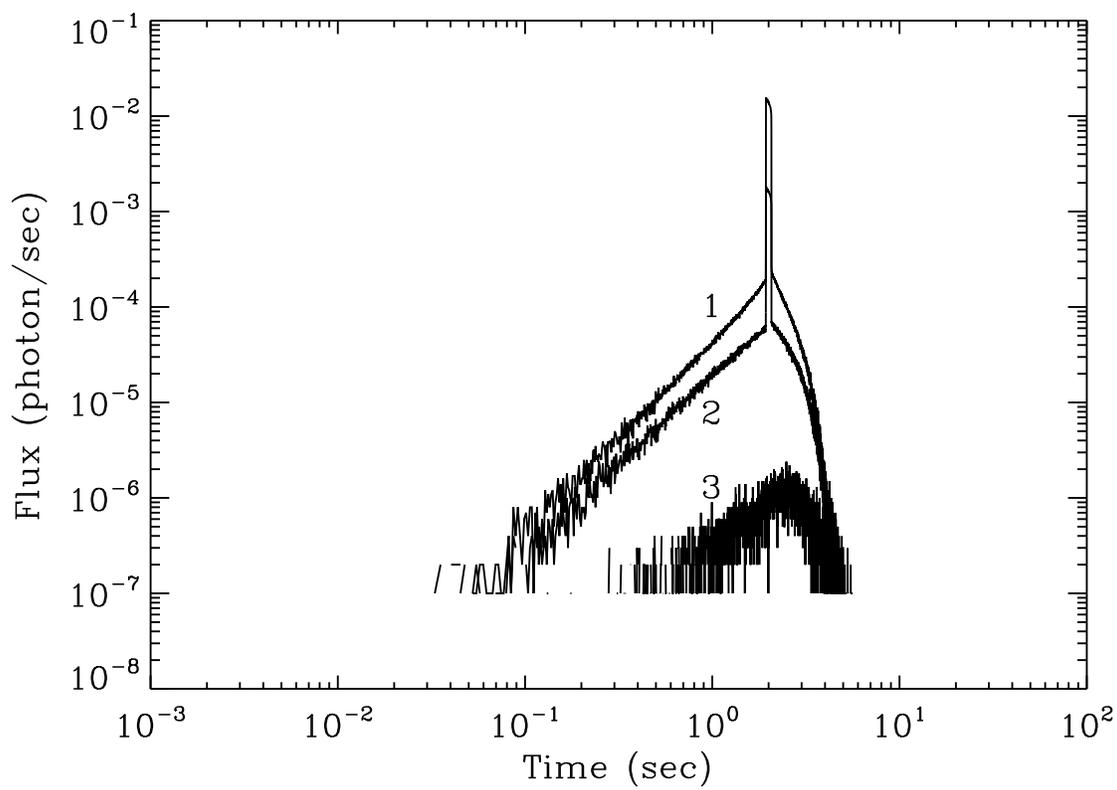
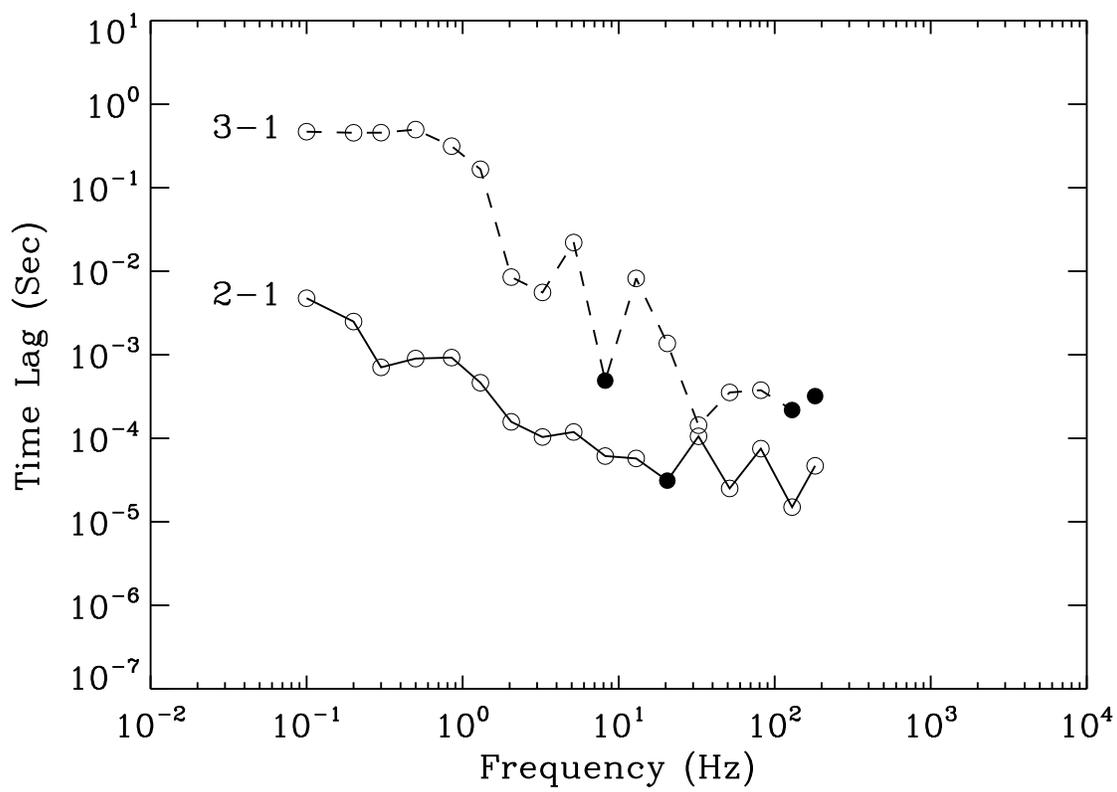
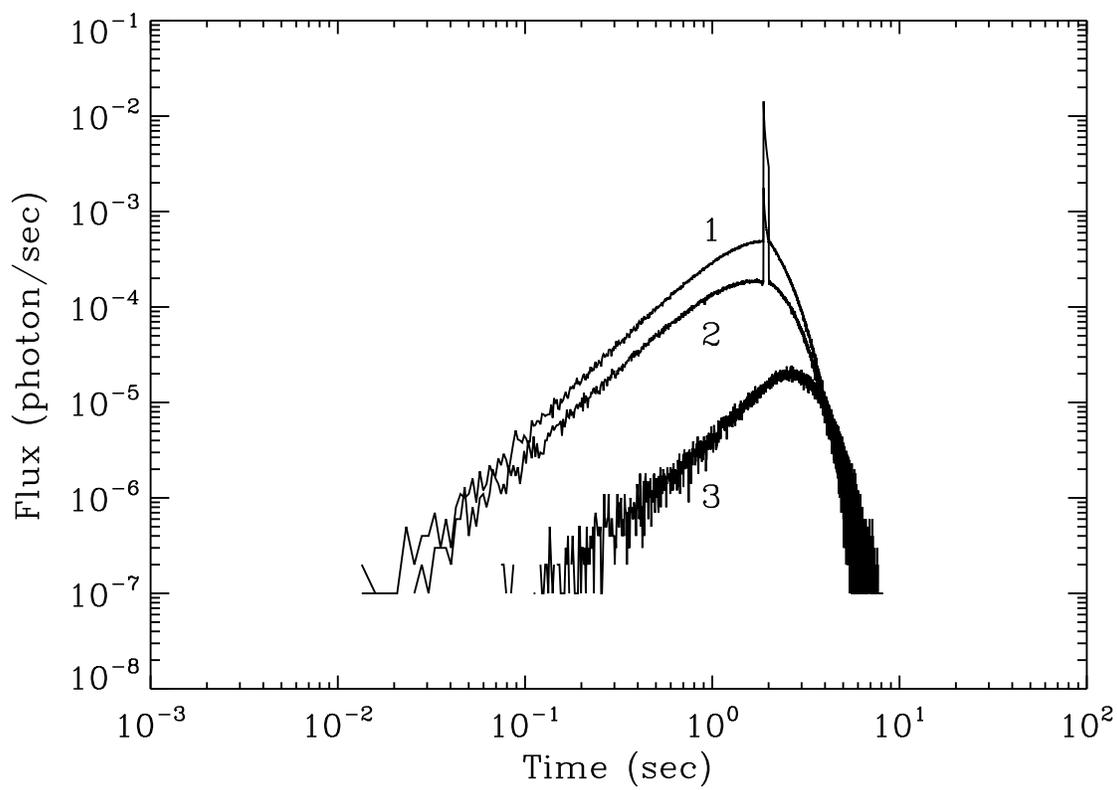
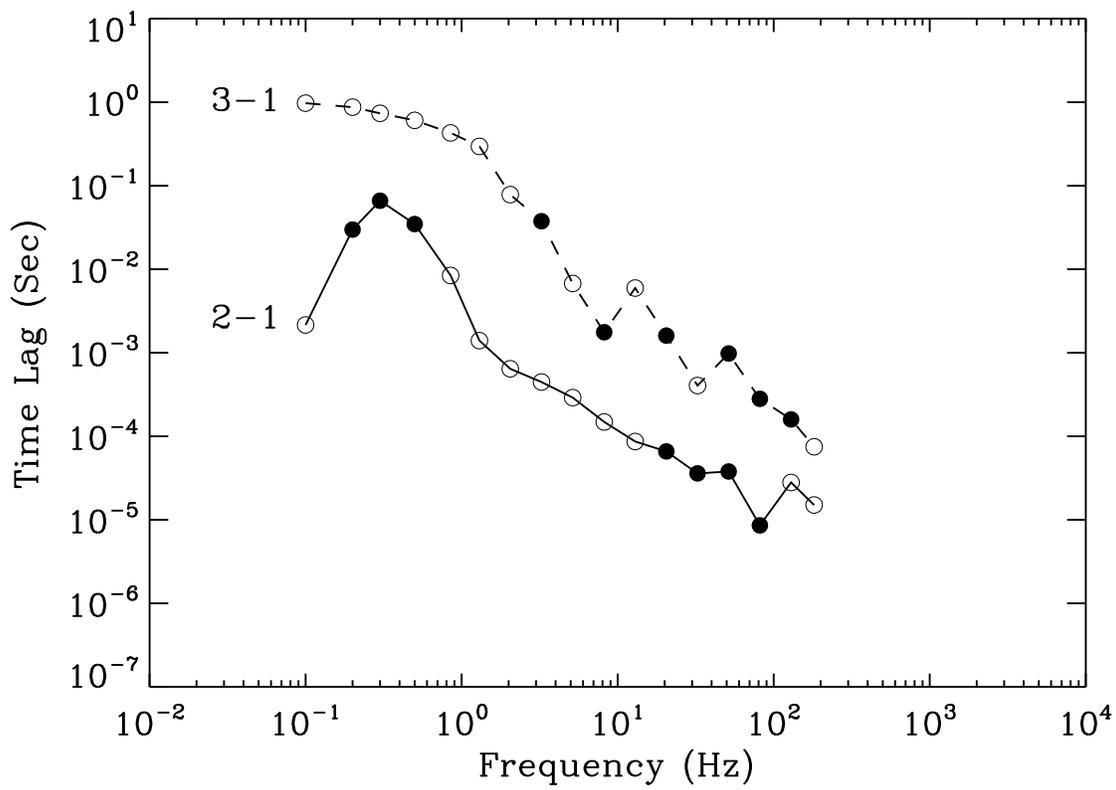
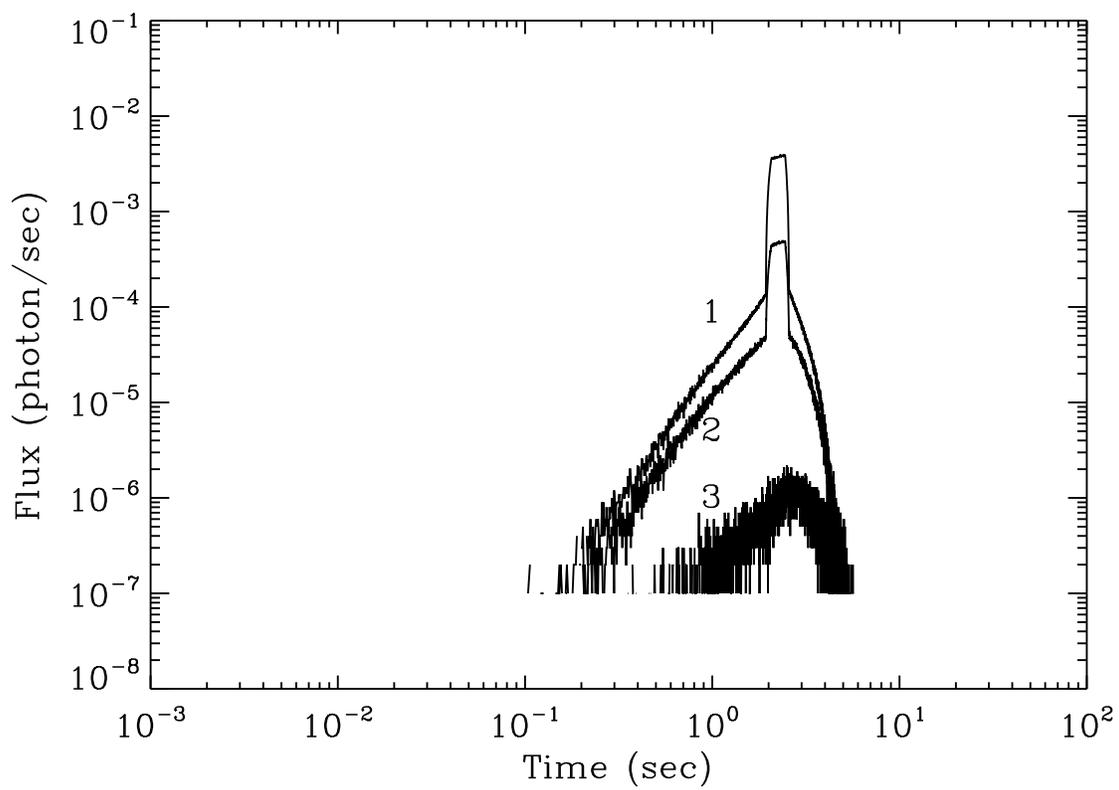
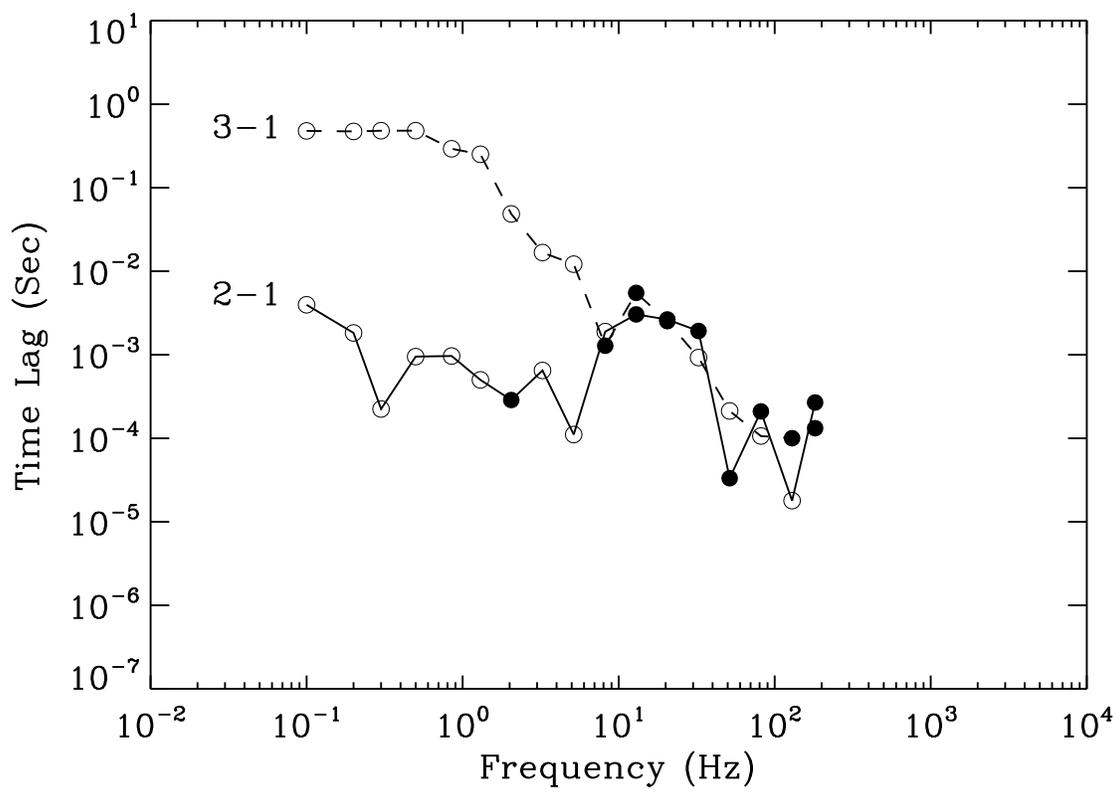
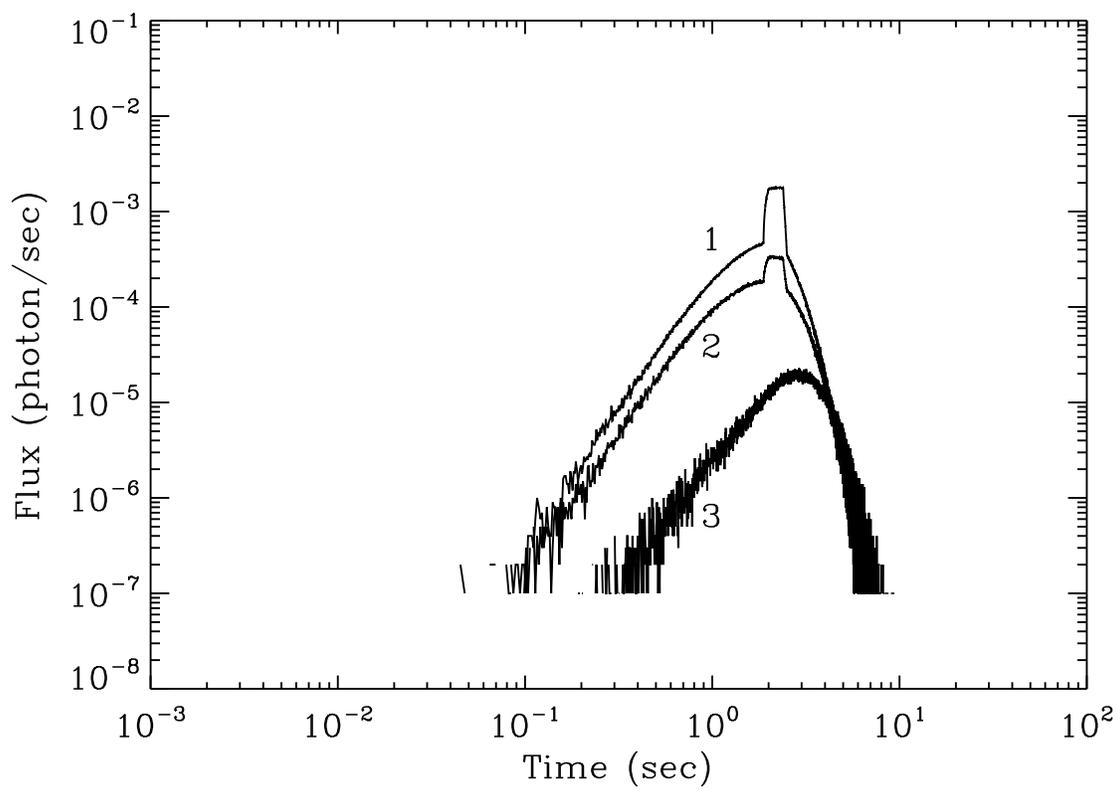
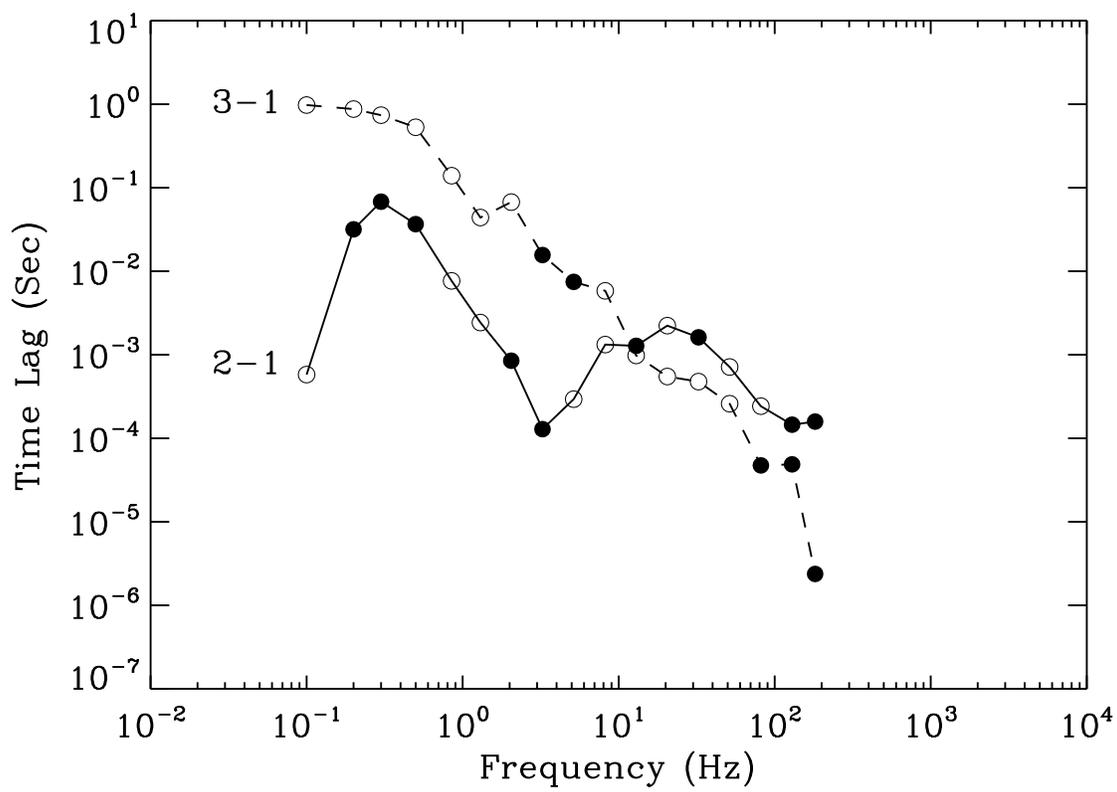
\end{document}